\documentclass{ewass_ss4proc}

\usepackage{kantlipsum}
\usepackage{seqsplit}

  \def \teff {$T_{\mathrm{eff}}$}
  \def \tc {$T_{\mathrm{c}}$}
  \def \vtur {$V_{\mathrm{tur}}$}
  \def \zettwo {$\zeta^2$ Ret}
  \def \zetone {$\zeta^1$ Ret}

\editors{Emeline Bolmont \& Sergi Blanco-Cuaresma}
\editorsurl{www.emelinebolmont.com | www.blancocuaresma.com/s/}
\publisher{Zenodo}
\conference{EWASS Special Session 4 (2017): Star-planet interactions}
\conferencedate{2017}

\title{The \tc \ trend in the $\zeta$ Reticuli system: $N$ spectra -- $N$ trends.}
\author{V. Adibekyan,$^{1}$
	P. Figueira,$^{1}$
	E. Delgado Mena,$^{1}$
	S. G. Sousa,$^{1}$
	N. C. Santos,$^{1,2}$ \\
        J.~I.~Gonz\'{a}lez Hern\'{a}ndez,$^{3,4}$
        G. Israelian$^{3,4}$}

\affiliation{$^{1}$ Instituto de Astrof\'isica e Ci\^encias do Espa\c{c}o, Universidade do Porto, CAUP, Rua das Estrelas, 4150-762 Porto, Portugal \\
			$^{2}$ Departamento de F\'isica e Astronomia, Faculdade de Ci\^encias, Universidade do Porto, Rua do Campo Alegre, 4169-007 Porto, Portugal \\
			  $^{3}$ Instituto de Astrof\'{\i}sica de Canarias, 38200 La Laguna, Tenerife, Spain \\
			   $^{4}$ Departamento de Astrof{\'\i}sica, Universidad de La Laguna, 38206 La Laguna, Tenerife, Spain}

\shorttitle{$\zeta$ Reticuli}
\shortauthors{Adibekyan et al.}

\abs{It is suggested that the chemical abundance trend with the condensation temperature, \tc, can be a signature of rocky planet formation or accretion.
Recently, a strong \tc \ trend was reported  in the   $\zeta$ Reticuli binary system \citep{Saffe-16}, where \zettwo \ shows a deficit of refractory elements relative to its companion (\zetone).
This depletion  was explained by the presence of a debris disk around \zettwo. Later, \citet{Adibekyan-16a} confirmed the significance of the trend, however, casted doubts on the interpretation proposed. 
Using three individual highest quality  spectra for each star, they  found that the \tc \ trends depend
on the individual spectra (three spectra of each star were used) used in the analaysis. In the current work we re-evaluated the presence and variability of the \tc \ trend in this system using a larger number of individual spectra.
In total, 62 spectra of \zettwo \ and 31 spectra of \zetone \ was used. Our results confirm the word of caution issued by \citet{Adibekyan-16a} that nonphysical factors can be
at the root of the \tc \ trends for the cases of individual spectra.}

\begin{document}

\maketitle

\section{Introduction}

Stars and planets form and evolve together. This common origin and evolution naturaly links some of the properties of stars and planets orbiting them.  There are many
vivid examples of such a bidirectional link between star-planet evolution: e.g. engulfment of planets due to stellar evolution \citep[e.g.][]{Kunitomo-11}, ejections of planets (leading to 
free floating planets) due to close fly-bys  \citep[e.g.][]{Hills-84}, white dwarfs atmospheric pollution due to acrretion of planets \citep[e.g.][]{Zuckerman-10},
and tidal evolution of close-in planets depending on the properties of the host star \citep[e.g.][]{Bolmont-17}.

Of all the various links and dependences between the properties of stars and planets, the dependence of planet formation and evolution on the chemical properties of the host star (and vice versa) is 
one of particular interest,  and is the subject of study of the current manuscript. Detection of only a few hot-Jupiters was enough to pinpoint the first correlation between
giant planet occurrence and the host star metallicity \citep[e.g.][]{Gonzalez-97, Santos-01}. This dependence, if it exists, is likely very weak for 
low-mass/small-size planets \citep[e.g.][]{Sousa-11}. The planet occurence -- metallicity correlation had a very important impact on the development of planet formation theories 
\citep[][]{Ida-04, Mordasini-12, Nayakshin-15}. Later, studies based on large data-sets showed that elements other than iron, such as, C, O, Mg, and Si, may also play a very 
important role for planet formation \citep[e.g.][]{Robinson-06, Haywood-09, Adibekyan-12a, Adibekyan-12b, Adibekyan-15, Adibekyan-17a}.

\begin{table}[t]
	
	\caption{The mean, standard deviation (of all the measurements) for stellar parameters and chemical abundances derived from individual
	spectra of \zetone \ and \zettwo.}
	\label{tab:table1}
	\begin{tabular*}{\linewidth}{l @{\extracolsep{\fill}}  l l l}
	\noalign{\smallskip}\hline\hline\noalign{\smallskip}
	 & \zetone & \zettwo \\
	\noalign{\smallskip}\hline\noalign{\smallskip}
	N spectra &  31 & 62 \\
	S/N &  3  218$\pm$54 &  327$\pm$41 \\
	\teff  & 5719$\pm$10  &  5851$\pm$9  \\
	$\log g$    &  4.51$\pm$0.02 &  4.48$\pm$0.02\\
	\vtur & 0.88$\pm$0.03   &  0.98$\pm$0.02 \\
	$[$Fe/H$]$   & -0.195$\pm$0.007  &  -0.206$\pm$0.007\\
	 \noalign{\smallskip}
	$[$NaI/H$]$ & -0.034$\pm$0.011 & -0.055$\pm$0.014  \\
	$[$MgI/H$]$ & -0.067$\pm$0.013  & -0.078$\pm$0.017  \\
	$[$AlI/H$]$ & -0.033$\pm$0.014  & -0.076$\pm$0.024   \\
	$[$SiI/H$]$ & -0.119$\pm$0.011   & -0.130$\pm$0.010  \\
	$[$CaI/H$]$ & -0.044$\pm$0.016   & -0.078$\pm$0.019   \\
	$[$ScII/H$]$ & -0.229$\pm$0.014  & -0.242$\pm$0.018  \\
	$[$<Ti>/H$]$ & -0.055$\pm$0.009   & -0.079$\pm$0.009  \\
	$[$VI/H$]$ & -0.173$\pm$0.013 & -0.222$\pm$0.013  \\
	$[$CrI/H$]$ & -0.135$\pm$0.009   & -0.172$\pm$0.011   \\
	$[$MnI/H$]$ & -0.242$\pm$0.023   & -0.303$\pm$0.018   \\
	$[$CoI/H$]$ & -0.181$\pm$0.012   & -0.21$\pm$0.0120   \\
	$[$NiI/H$]$ & -0.199$\pm$0.007  & -0.215$\pm$0.008  \\
	\noalign{\smallskip}\hline
	\end{tabular*}
\end{table}

\begin{figure*}
\begin{center}$
\begin{tabular}{cc}
\includegraphics[width=0.46\linewidth]{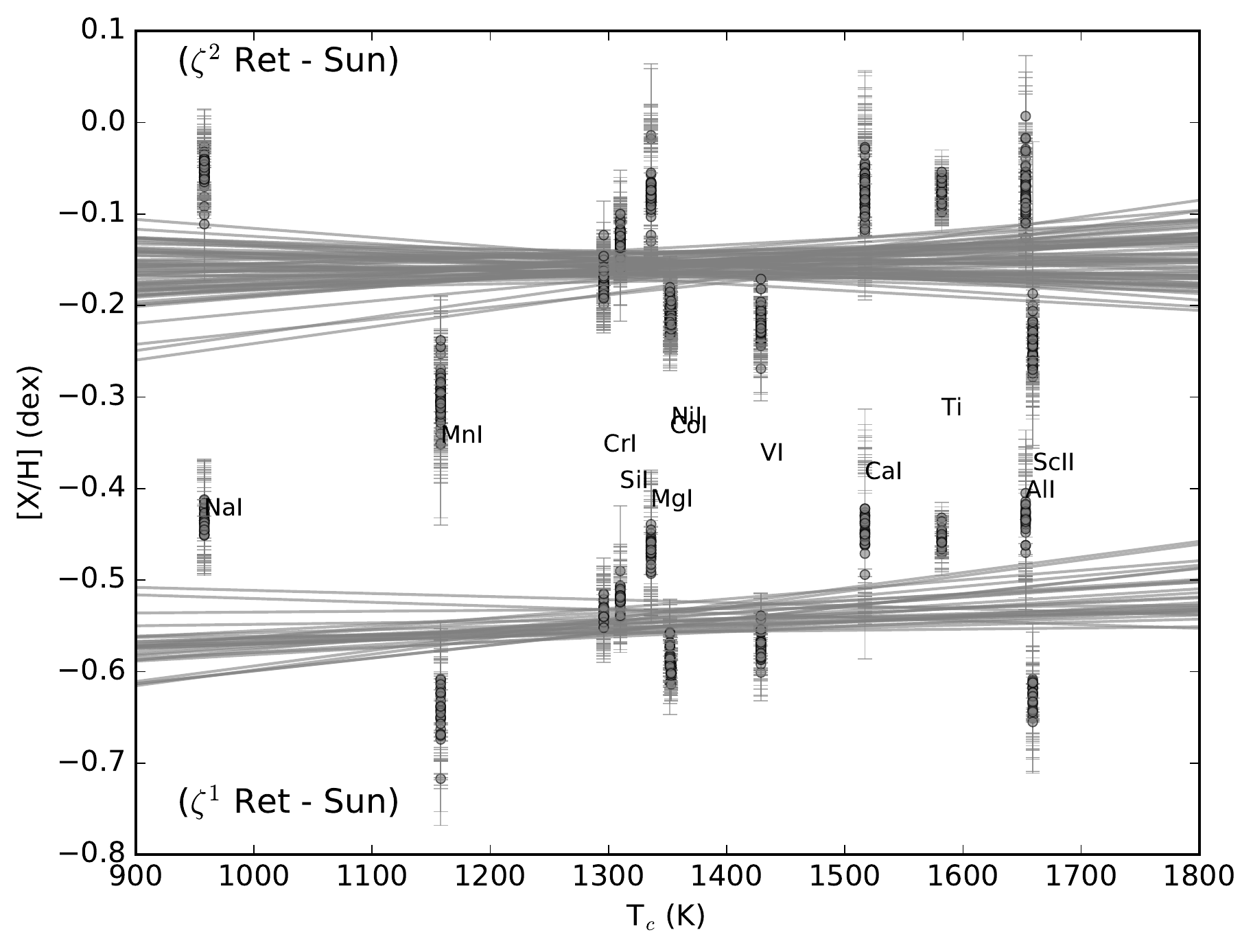}&
\includegraphics[width=0.45\linewidth]{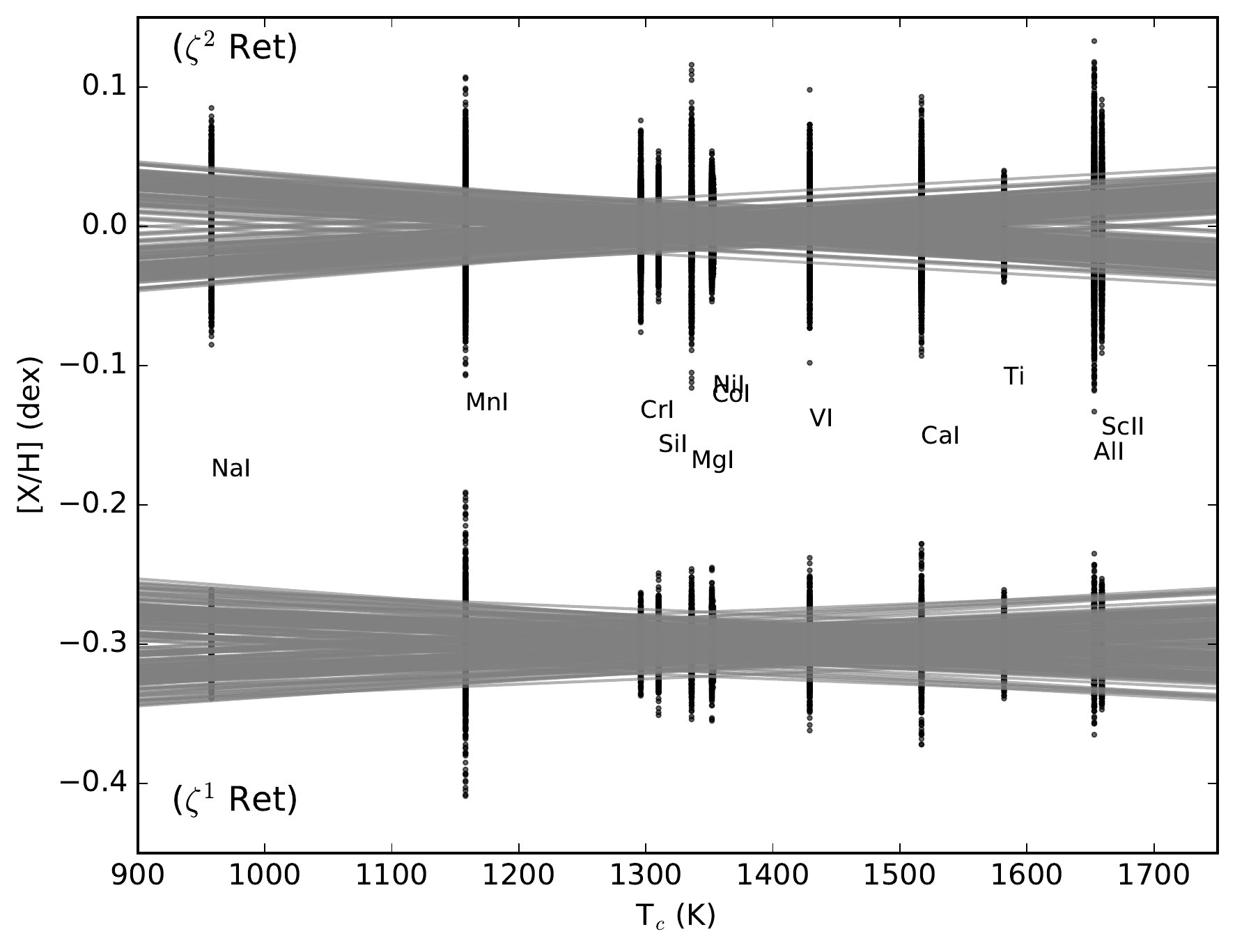}
\end{tabular}$
\end{center}
\vspace{-0.8cm}
\caption{\textit{Left:} [X/H] against condensation temperature  for individual spectra of \zetone \ and \zettwo \ relative to the Sun.  The gray lines represent
the results of  the linear regression. \textit{Right:} Abundances against condensation temperature for \zetone \ and \zettwo, derived from each individual spectra relative to the other 
individual spectra of the same star. The gray lines show the results of the linear regression of the 100 steppes negative and positive trends.  The results for \zetone \ are offseted by
-0.3 dex for a sake of visibility.}
\label{fig1}
\end{figure*}

Many astronomers, starting from \citet{Gonzalez-97} and \citet{Smith-01}, also tried to search for chemical signatures of planet formation and planet accretion on the planet-host stars. 
In particular, the presence of a trend between the abundances of chemical elements and the condensation temperature of the elements was explored. 
This trend is usually called \tc \ trend, and the slope of the correlation  of [X/Fe] vs. condensation temperature is usually named \tc \ slope.

\citet{Melendez-09} were the first to report a statistically significant deficit of refractory elements (high-\tc) with respect to volatiles (low-\tc) in the Sun when
compared to solar twin stars. Together with the rocky material accretion \citep[e.g.][]{Schuler-11} and/or rocky material trap  in 
terrestrial planets \citep[e.g.][]{Melendez-09}, several explanations are proposed to explain the \tc \ trend. \citet{Adibekyan-14} suggested that the \tc \ trend strongly depends on the stellar age 
and they found a tentative dependence on the galactocentric distances of the stars. The correlation with stellar age was later confirmed by several authors 
\citep[e.g.][]{Nissen-15, Spina-16}, while the possible relation with the galactocentric distances seems to be  more challenging \citep[see][]{Adibekyan-16b}.
\citet{Maldonado-15} and \citet{Maldonado-16} further suggested a significant correlation with the stellar radius and mass.
This very exciting possible  connection between chemical peculiarities of parent stars and formation of planets has also been examined in 
other works \citep[e.g.][]{Ramirez-09,  Biazzo-15, Saffe-16, Mishenina-16, GH-10, GH-13, Teske-16}, 
but contradictory conclusions were reached. For more discussion and references we refer the reader to \citet{Adibekyan-17b}.

Very recently, \citet[][]{Saffe-16} reported a positive \tc \ trend in the binary system, \zetone \ -- \zettwo, \ where one of the stars (\zettwo) hosts a debris disk. 
The authors explained the deficit of the refractory elements relative to volatiles in \zettwo\  as caused by the depletion of about $\sim$3 M$_{\oplus}$  rocky material.
In the following study, \citet{Adibekyan-16a} confirmed the trend obtained by \citet{Saffe-16} using higher S/N data. However, they also found that 
non-astrophysical factors, such as the quality of spectra employed and errors that are not accounted for, can be responsible for the \tc \ trends. When using the three highest 
signal-to-noise (S/N) spectra of each component of the binary system, the authors found significant but varying differences in the abundances of the same star from different individual spectra. 

In this work we used larger number of spectra of  \zetone \ and \zettwo \ to expand on the analysis and perform a double-check on the results obtained in \citet{Adibekyan-16a}.

\section{Data and analysis}

The HARPS archive (ESO archive phase 3) contains $\sim$70 and $\sim$170 spectra for \zetone \ and \zettwo, respectively. From this archive we selected the highest quality spectra
for both stars: 31 spectra with S/N $>$ 150 for \zetone \ and 62 spectra with S/N $>$ 250 for \zettwo \ (see Table\,\ref{tab:table1}). Then for each individual spectra we derived stellar 
parameters and chemical abundances. Stellar parameters were  derived as in our previous works \citep[e.g.][]{Sousa-14, Andreasen-17}. 
We first automatically measured the equivalent widths  using the ARES v2 code\footnote{The last version of ARES code (ARES v2) can 
be downloaded at http://www.astro.up.pt/$\sim$sousasag/ares} \citep{Sousa-15} and then used the grid of ATLAS9 plane-parallel model
of atmospheres \citep{Kurucz-93} and the 2014 version of MOOG\footnote{The source code of MOOG can be downloaded at
http://www.as.utexas.edu/$\sim$chris/moog.html}  radiative transfer code \citep{Sneden-73} to derive parameters under assumption of LTE. The same tools were used to derive chemical abundances of 12
elements following the procedure described in our previous works \citep[e.g.][]{Adibekyan-12c, Adibekyan-15a, Adibekyan-15b}.
We note that the parameters and abundances are derived  with a classical rather than a line-by-line differential approach. In Table\,\ref{tab:table1} we present the mean and 
standard deviation of obtained values for stellar parameters and chemical abundances of each star. The table shows that in average the spectra-to-spectra 
dispersion  for all the parameters and abundances is very small. However, for individual cases some of the parameters or abundances can be different from the mean value and may 
have a significant impact on the derived \tc \ trend.

\begin{figure}[ht]
	\centering
	\includegraphics[width=1.0\linewidth]{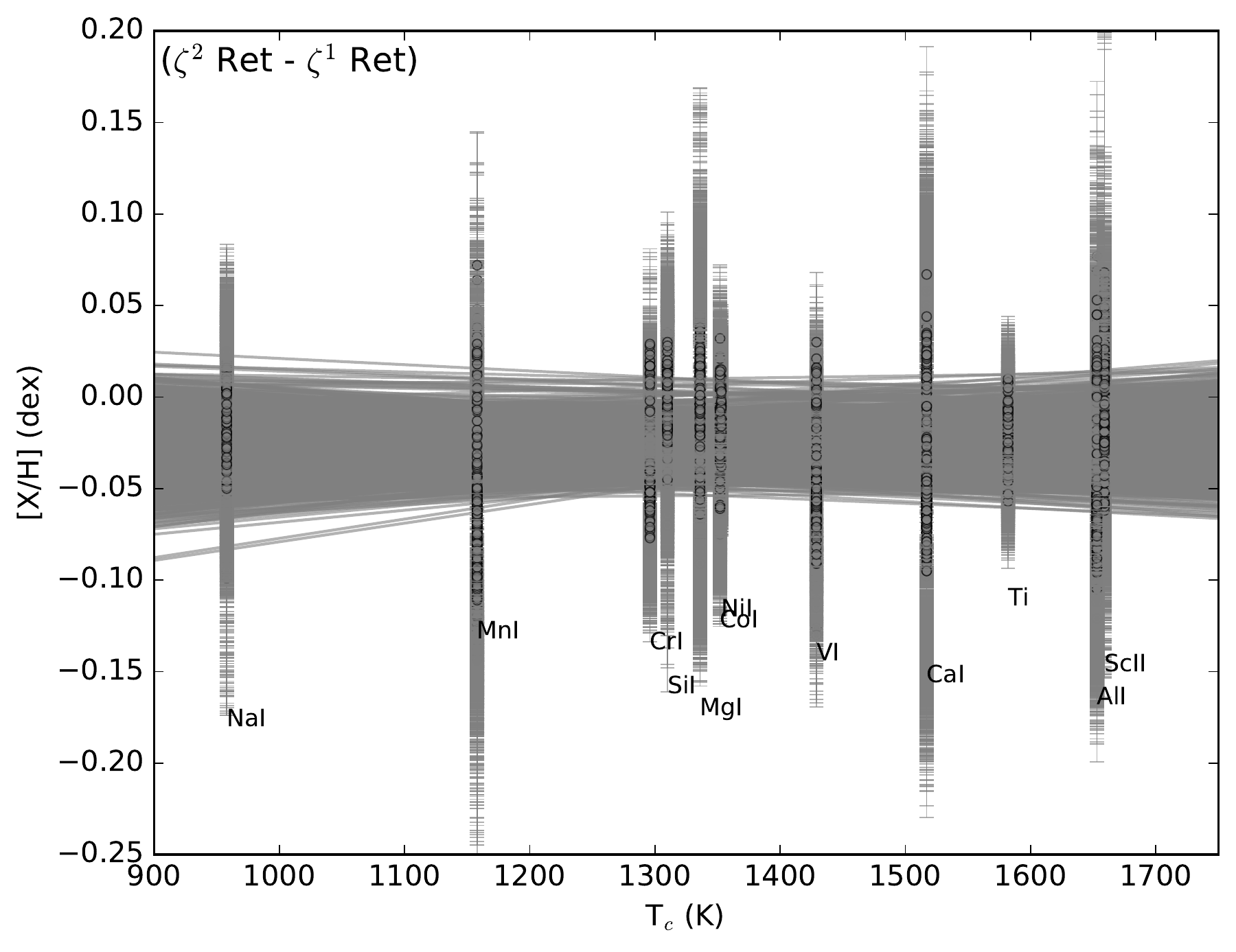}
	\vspace{-.7cm}
	\caption{[X/H] against condensation temperature  for individual spectra of \zettwo \ relative to \zetone. The gray lines represent the results of  the linear regression.}
	\label{fig2}
\end{figure}

\begin{table*}[t]
	\begin{center}
	\caption{Slopes of the [X/H] vs. \tc \ for different pairs of spectra when the most significant positive and negative trends are observed. A frequentist approach is chosen to derive 
the slopes and their uncertainties.}
	\label{tab:table2}
	\begin{tabular*}{\linewidth}{l @{\extracolsep{\fill}} c l}
	\noalign{\smallskip}\hline\hline\noalign{\smallskip}
	Star (S/N: spectra) & {Slope$\pm \sigma$} & P(F-stat) \\
	\noalign{\smallskip}\hline\noalign{\smallskip}
	\zettwo \ (278: HARPS.2010-02-21T00:11:27.058) - \zettwo  \ (252: HARPS.2008-10-07T05:22:11.508) & -8.24 $\pm$  3.47 & 0.039 \\
	\zettwo \ (268: HARPS.2008-10-07T05:31:04.470) - \zettwo  \ (292: HARPS.2009-12-01T01:32:07.799) & 7.59 $\pm$  3.40 & 0.050 \\
	\zetone \ (174: HARPS.2007-09-07T09:09:24.860) - \zetone  \ (169: HARPS.2004-10-28T07:52:55.573) & 6.39 $\pm$  2.24 & 0.017 \\
	\zetone \ (169: HARPS.2004-10-28T07:52:55.573) - \zetone  \ (166: HARPS.2005-09-11T09:45:41.214) & -6.86 $\pm$  1.65 & 0.002 \\
	\zettwo \  (268: HARPS.2008-10-07T05:31:04.470) - \zetone  \ (160: HARPS.2007-07-29T10:11:20.240) & 8.49 $\pm$  4.02 & 0.061 \\
	\zettwo \  (359: HARPS.2008-11-30T03:42:06.702) - \zetone  \ (369: HARPS.2005-11-15T03:52:37.899) & -6.65 $\pm$  3.88 & 0.117 \\
	\noalign{\smallskip}\hline
	\end{tabular*}
	\end{center}
	\vspace{-0.2cm}
	Note: Units of the slopes are in 10$^{-5}$ dex K$^{-1}$.
\end{table*}

\section{Results}

\subsection{\zetone \ and  \zettwo \ vs. Sun}

On the left panel of  Fig.~\ref{fig1} we show the dependence of [X/H] abundances of \zetone \ and \zettwo \ relative to the Sun on the corresponding \tc. 
The 50\% \tc \ equilibrium condensation temperatures for a gas of solar system composition are taken from \citet{Lodders-03}. 
We calculated the slopes with the weighted least-squares (WLS) technique, whereas we calculated the weights as the inverse of the variance ($\sigma^2$) of the abundance. The p-values come from 
the F-statistics that tests the null hypothesis that the data can be modeled accurately by setting the regression coefficients to zero.
We refer the reader to \citet{Adibekyan-16a} for more details. From the figure one can clearly see that depending on the individual spectra the trend can be negative or positive. However, we should note
that none of the trends are statistically significant (i.e. different from zero) with the lowest P-value being $\sim$0.17. Interestingly, using  very high S/N combined spectra of \zettwo, \citet{Adibekyan-16a} obtained 
a P-value (using the same F statistics) of $\sim$0.03 when many elements of wide range of \tc \ were used. When only elements with \tc \ $>$ 900 K were used the authors obtained a p-value of 0.7
(i.e., no statitical evidence for the difference from no slope).

\subsection{\zetone \ vs. \zetone \ and  \zettwo \ vs. \zettwo}

On the right panel of  Fig.~\ref{fig1} we show the dependence of [X/H] abundances against corresponding \tc \ for \zetone \ and \zettwo \ derived for each individual spectra
using all other individual spectra as reference. Again, as for the abundances relative to the Sun, we can see that the sign and significance of the trend depends on the individual spectra 
used. In Table\,\ref{tab:table2} we provide the results for the most significant trends. The table shows that both negative and positive significant trends can be obtained for each star when
individual spectra are used. This result is in good agreement with those obtained in \citet{Adibekyan-16a}.

\subsection{\zetone \ vs.  \zettwo}

Following the same logic, in  Fig.~\ref{fig2} we show the dependence of [X/H] abundances against corresponding \tc \ for individual spectra of \zettwo \ relative to \zetone.
The results are not different from what was obtained in the previous two subsections: combination of different individual spectra lead to different results.
We should remind that when the combined higher S/N spectra were used, both \citet{Saffe-16} and \citet{Adibekyan-16a} found a significant abundance difference between \zettwo \ and \zetone \
that correlates with \tc. We will not compre the exact values of the slopes otained in the current work and in \citet{Saffe-16} and \citet{Adibekyan-16a}, because different number of
chemical elements (with a different range of condensation temperature) were used in these works.

\section{Conclusion}
The $\zeta$ Reticuli binary system is composed of two solar analogs where one of the components (\zettwo) hosts a debris disk. \citet{Saffe-16} found that the component hosting the debris
shows a deficit of refractory-to-volatile elements when compared to its companion. The authors showed that the abundance difference between the two components correlate with the \tc.
Later, \citet{Adibekyan-16a} confirmed this result but also found that the trends (and their significance) obtained in these works depend on the spectra used.

We used 31 high-quality (S/N $>$ 150) individual spectra of \zetone \ and 62 high-quality (S/N $>$ 250) individual spectra of \zettwo \  to revisit the results 
obtained in \citet{Adibekyan-16a}, namely that the \tc \ trend depends on the individual spectra used. 

Our results show that indeed the \tc \ trend depends on the spectra used in the analysis and thus show that there are nonastrophysical factors 
that may be responsible for the observed \tc \ trends. These results also indicate that one should be careful when analyzing very subtle differences in chemical abundances between stars. 
When studying  the chemical abundance trends with condensation temperature, it is very important to use very high-quality combined spectra.
The combination of many spectra increase the S/N and may minimize possible time-dependent effects.

\section*{Acknowledgments}

{V.A. thanks the organizers of EWASS Special Session 4 (2017), Emeline Bolmont \& Sergi Blanco-Cuaresma, for a very interesting session and for selecting his oral contribution. 
This work was supported by Funda\c{c}\~ao para a Ci\^encia e Tecnologia (FCT) through national funds (ref. PTDC/FIS-AST/7073/2014
and ref. PTDC/FIS-AST/1526/2014) through national funds and by FEDER through COMPETE2020 (ref. POCI-01-0145-FEDER-016880 and ref. POCI-01-0145-FEDER-016886).
V.A., E.D.M., P.F., N.C.S., and S.G.S. also acknowledge the support from FCT through Investigador FCT contracts of reference \seqsplit{IF/00650/2015/CP1273/CT0001,} \seqsplit{IF/00849/2015/CP1273/CT0003,} 
\seqsplit{IF/01037/2013/CP1191/CT0001,} \seqsplit{IF/00169/2012/CP0150/CT0002,} and \seqsplit{IF/00028/2014/CP1215/CT0002}, respectively, and POPH/FSE (EC) by FEDER funding through 
the program ``Programa Operacional de Factores de Competitividade - COMPETE''. 
PF further acknowledges support from FCT in the form of an exploratory project of reference IF/01037/2013CP1191/CT0001. 
G.I. acknowledges financial support from the Spanish Ministry project MINECO AYA2011-29060.
JIGH acknowledges financial support from the Spanish Ministry of Economy and Competitiveness (MINECO) 
under the 2013 Ram\'{o}n y Cajal program MINECO RYC-2013-14875, and the Spanish ministry project MINECO AYA2014-56359-P.}

\bibliographystyle{ewass_ss4proc}
\bibliography{references.bib}

\end{document}